# Phase-Controlled Atom-Photon Entanglement in a Three-Level Λ-Type Closed-Loop Atomic System

**Ali Mortezapour**[a)], **Zeinab Kordi**[b)] **and Mohammad Mahmoudi**[b)†]

[a)] Institute for Advanced Studies in Basic Sciences, P.O. Box 45195-159, Zanjan, Iran

[b)] Physics Department, University of Zanjan, P.O. Box 45195-313, Zanjan, Iran

[†]Email: mahmoudi@znu.ac.ir

Abstract. We study the entanglement of dressed atom and its spontaneous emission in a three-level Λ-type closed-loop atomic system in multi-photon resonance condition and beyond it. It is shown that the von Neumann entropy in such a system is phase dependent, and it can be controlled by either intensity or relative phase of applied fields. It is demonstrated that for the special case of Rabi frequency of applied fields, the system is disentangled. In addition, we take into account the effect of Doppler broadening on the entanglement and it is found that a suitable choice of laser propagation directions allows us to obtain the steady state degree of entanglement (DEM) even in the presence of Doppler effect.

## 1. Introduction

Entanglement is one of the most intriguing features of quantum mechanics, which emanates from quantum correlation between different parts of a system. The concept of entanglement and the name go back to the famous EPR article and Schrödinger's rejoinder [1] in 1935 on the foundations of quantum measurement. Generally, a system consisting of two sub-system is said to be entangled if its quantum state cannot be described by a simple product of the quantum states of the two components. [2] One consequence of this form of the wave function is that a measurement on one of the sub-system yields the information about **the** other sub-system.

Recently, it has become clear that the importance of the phenomenon extends well beyond fundamental questions about quantum theory; indeed, entanglement lies at the heart of most of the quantum-information processing tasks such as quantum computing[3], quantum teleportation[4,5] and quantum cryptography,[6,7] dense coding.[8]

Entanglement can occur as a result of interaction between different parts of a system. One of the fundamental interactions in physics is the interaction between light and matter. In the recent years theoretical description of entanglement evolution between atom and quantized field in the Jaynes-Cummings model has been proposed.[9-15] Moreover, entanglement of an atom and its spontaneous emission was investigated.[16] The effect of spontaneously generated coherence (SGC) on the atom-photon entanglement was also studied.[17] It was also shown that the induced steady state entanglement between atom and its spontaneous emission can be controlled by the intensity and detuning of applied field.[18] In another study, it was shown that the atom-photon entanglement depends on the relative phase of applied fields.[19]

Generally, atom-photon entanglement can be used in new interesting quantum concepts. "Many new concepts, for example, the quantum repeater .[20]or quantum networks for distributed quantum computing thus require the faithful mapping of quantum information between photonic quantum channels and matter-based quantum memories and processors. Entanglement between matter and light is crucial for achieving this task."[21-22]

It has been demonstrated that the optical properties of quantum system containing the closed-loop configuration in multi-photon resonance condition are phase-dependent.[23-25] In this paper, we investigate the entanglement of dressed atom and its spontaneous emission in a three-level Λ-type closed-loop atomic system. It is shown that the entanglement can be controlled by either intensity or relative phase of applied fields. In addition, we found that under the special condition the atom and photons become disentangled. Note that the phase-dependent behavior, here, arise from the closed-loop configuration but in our previous work[19] it has been applied by the spontaneously generated coherence of spontaneous emission.[26]

**2. Model and Solution**

Consider a Λ-type three-level atomic system including two ground state $|1\rangle$, $|2\rangle$ and an excited-state $|3\rangle$ which is coupled by two coherent optical fields $\vec{E}_R$, $\vec{E}_L$ and a microwave driving field $\vec{E}_m$ as shown in figure 1. The left(right) field $\vec{E}_L = E_L \exp[-i(\omega_L t + \varphi_L)] + h.c$ ($\vec{E}_R = E_R \exp[-i(\omega_R t + \varphi_R)] + h.c$) of frequency $\omega_L$ ($\omega_R$) and initial phase $\varphi_L$ ($\varphi_R$) drives the transition $|1\rangle \leftrightarrow |3\rangle$ ($|2\rangle \leftrightarrow |3\rangle$) with the Rabi frequency $\Omega_L = \vec{E}_L.\vec{d}_{13}/\hbar$ ($\Omega_R = \vec{E}_R.\vec{d}_{23}/\hbar$), while the transition $|1\rangle \leftrightarrow |2\rangle$ is excited by a microwave driving field $\vec{E}_m = E_m \exp[-i(\omega_m t + \varphi_m)] + h.c$ of frequency $\omega_m$, with initial phase $\varphi_m$ and Rabi frequency $\Omega_m = \vec{E}_m.\vec{d}_{12}/\hbar$. It is well-known that existence of microwave driving field

renders this Λ-type system into a closed-loop system. Here $E_L(E_R)$ and $E_m$ are the amplitudes of the left (right) and microwave driving fields, respectively. $\vec{d}_{12}$, $\vec{d}_{13}$ and $\vec{d}_{23}$ display the corresponding dipole moments of transitions. The spontaneous decay rates from $|3\rangle$ to level $|1\rangle$ ($|2\rangle$) is denoted by $2\gamma_{31}$ ($2\gamma_{32}$) while the spontaneous decay from level $|2\rangle$ to level $|1\rangle$ has been ignored.

Under the dipole and the rotating wave approximation, the total Hamiltonian of this system in the interaction picture reads:

$$H = -\hbar\Omega_L e^{-i(\Delta_L t + \varphi_L)}|3\rangle\langle 1| - \hbar\Omega_R e^{-i(\Delta_R t + \varphi_R)}|3\rangle\langle 2| - \hbar\Omega_m e^{-i(\Delta_m t + \varphi_m)}|2\rangle\langle 1| + H.c., \quad (1)$$

where $\Delta_L = \omega_L - \omega_{31}$, $\Delta_R = \omega_R - \omega_{32}$ and $\Delta_m = \omega_m - \omega_{21}$ are the detunings between applied fields and the corresponding atomic transition frequencies. $\omega_{21}$, $\omega_{31}$ and $\omega_{32}$ stand for corresponding frequency of transitions $|1\rangle \leftrightarrow |2\rangle$, $|1\rangle \leftrightarrow |3\rangle$ and $|2\rangle \leftrightarrow |3\rangle$, respectively. The so-called three-photon detuning and relative phase are defined as $\Delta = \Delta_L - \Delta_R - \Delta_m$ and $\varphi = \varphi_L - \varphi_R - \varphi_m$, respectively. For simplicity we set $\hbar = 1$ and take the Rabi frequencies of three classical fields as real.

The density matrix equations of motion for this system can be written as:

$$\dot{\rho}_{11} = i\Omega_m \rho_{21} e^{-i\varphi} + i\Omega_L^* \rho_{31} e^{i\Delta t} - i\Omega_m \rho_{12} e^{i\varphi} - i\Omega_L \rho_{13} e^{-i\Delta t} + 2\gamma_{31}\rho_{33},$$

$$\dot{\rho}_{22} = i\Omega_m \rho_{12} e^{i\varphi} + i\Omega_R \rho_{32} - i\Omega_m \rho_{21} e^{-i\varphi} - i\Omega_R \rho_{23} + 2\gamma_{32}\rho_{33},$$

$$\dot{\rho}_{12} = -i\Delta_m \rho_{12} + i\Omega_m(\rho_{22} - \rho_{11})e^{-i\varphi} + i\Omega_L^* \rho_{32} e^{i\Delta t} - i\Omega_R \rho_{13},$$

$$\dot{\rho}_{13} = -i(\Delta_L - \Delta)\rho_{13} + i\Omega_L^*(\rho_{33} - \rho_{11})e^{i\Delta t} + i\Omega_m \rho_{23} e^{-i\varphi} - i\Omega_R \rho_{12} - (\gamma_{31} + \gamma_{32})\rho_{13},$$

$$\dot{\rho}_{23} = -i\Delta_R \rho_{23} + i\Omega_m \rho_{13} e^{i\varphi} + i\Omega_R(\rho_{33} - \rho_{22}) - i\Omega_L^* \rho_{21} e^{i\Delta t} - (\gamma_{31} + \gamma_{32})\rho_{23},$$

$$\rho_{11}(t) + \rho_{22}(t) + \rho_{33}(t) = 1. \quad (2)$$

Now, we seek the corresponding steady state analytical solution for elements of the density matrix for $\gamma_{31} = \gamma_{32} = \gamma = 1$ and $\Omega_R = \Omega_L = \Omega_0$. The population and coherence terms of the density matrix for $\Delta_L = \Delta_R = \Delta_m = 0.0$ are given by:

$$\rho_{11} = \frac{\Omega_m^4 + \Omega_0^4 - \Omega_m^2(-4 + \Omega_0^2) - \Omega_m\Omega_0^2(\Omega_m Cos[2\varphi] + 2Sin[\varphi])}{2\Omega_m^4 + 2\Omega_0^4 - \Omega_m^2(-8 + \Omega_0^2) - 3\Omega_m^2\Omega_0^2 Cos[2\varphi]},$$

$$\rho_{22} = \frac{\Omega_m^4 + \Omega_0^4 - \Omega_m^2(-4 + \Omega_0^2) + \Omega_m\Omega_0^2(-\Omega_m Cos[2\varphi] + 2Sin[\varphi])}{2\Omega_m^4 + 2\Omega_0^4 - \Omega_m^2(-8 + \Omega_0^2) - 3\Omega_m^2\Omega_0^2 Cos[2\varphi]},$$

$$\rho_{12} = \frac{(e^{-2i\varphi}\Omega_m^2(-4 - \Omega_m^2 + \Omega_0^2) - (\Omega_m^4 + 2\Omega_0^4 + \Omega_m^2(4 - 3\Omega_0^2)))}{2(2\Omega_m^4 + 2\Omega_0^4 - \Omega_m^2(-8 + \Omega_0^2) - 3\Omega_m^2\Omega_0^2 Cos[2\varphi])},$$

$$\rho_{13} = \frac{(1-e^{-2i\varphi})\Omega_m\Omega_0(2i\Omega_m + e^{i\varphi}(\Omega_m^2 - \Omega_0^2))}{-2(2\Omega_m^4 + 2\Omega_0^4 - \Omega_m^2(-8+\Omega_0^2) - 3\Omega_m^2\Omega_0^2 Cos[2\varphi])},$$

$$\rho_{23} = \frac{2iSin[\varphi]\Omega_m\Omega_0(\Omega_0^2 - \Omega_m(\Omega_m + 2iCos[\varphi] - 2Sin[\varphi]))}{-2(2\Omega_m^4 + 2\Omega_0^4 - \Omega_m^2(-8+\Omega_0^2) - 3\Omega_m^2\Omega_0^2 Cos[2\varphi])},$$

$$\rho_{11} + \rho_{22} + \rho_{33} = 1. \tag{3}$$

It is clear that all expressions in Equation (3) are phase-dependent.

## 3. The Evolution of Entropy and Atom-photon Entanglement

Think over a bipartite quantum system described by a density operator $\rho_{AF}$ in a tensor product space $C_A \otimes C_F$. $C_{A(F)}$ is the Hilbert space of the subsystem $A(F)$. The partial density operator of one part will be obtained by tracing over the other:

$$\rho_{A(F)} = Tr_{F(A)}(\rho_{AF}). \tag{4}$$

This bipartite quantum system is called separable, if its density operator can be written as

$$\rho_{AB} = \rho_A \otimes \rho_B, \tag{5}$$

Otherwise it is called an entangled state.[27]

In addition to the generation of entangled states, measurement of the entanglement in general composite systems is a more interesting problem of researchers at the moment. Here we employ the von Neumann entropy $S$ as a unique measure to estimate the amount of atom-field entanglement in our system. For a system in quantum state $\rho$, this unique measure is defined as [28]:

$$S = -Tr\,\rho\ln\rho, \tag{6}$$

Araki and Lieb in their fantastic paper [29] show that for a bipartite quantum system composed of two subsystems A and F (say the atom and field) at any time t, the system and subsystem's entropies satisfy a remarkable inequality as bellow:

$$|S_A(t) - S_F(t)| \le S_{AF}(t) \le |S_A(t) + S_F(t)|, \tag{7}$$

Where $S_{AF}$ is the total entropy of the composite system and $S_{A(F)}(t) = -Tr[\rho_{A(F)}(t)\ln\rho_{A(F)}(t)]$ are partial entropies corresponding to reduced density operators.

Based on Equation (7), for a closed atom–field system in which both of them start from pure state, the entropies of two interacting subsystems will be precisely equal at all times after the interaction of two subsystems are switched on. Then Phoenix and Knight [30, 31] have shown that under the aforementioned circumstances, the information about any of the subsystems is an indication of the entanglement of the whole system, namely a decrease in partial entropy means that each subsystem evolves towards a pure quantum state, whereas a rise in partial entropy means that the two components tend to lose their individuality and become correlated or entangled. The degree of entanglement (DEM) for atom-field entanglement is defined as:

$$DEM(t) = S_A = S_F = -(\sum_{j=1}^{3} \lambda_j \ln \lambda_j), \qquad (8)$$

where $\lambda_j$ is the eigenvalue of the reduced density matrix of the atom.

**4- Results and discussion**

In this section, we depict our results for the behavior of the system in three-photon resonance condition and beyond it. All parameters are reduced to dimensionless units through scaling by $\gamma_{31} = \gamma_{32} = \gamma$ and all figures are plotted in the unit of $\gamma$. We first discuss the dynamical behavior of the reduced atomic entropy via different values of intensity of applied fields.

In figure 2, we plot the dynamical behavior of DEM in multi-photon resonance condition. Using parameters are $\Delta_L = \Delta_R = \Delta_m = 0.0\gamma$, $\Omega_L = 0.25\gamma$, $\Omega_R = 0.25\gamma$ (solid), $0.5\gamma$ (dashed), $\Omega_m = 0$ (a), $0.5\gamma$ (b). An investigation on figure 2 shows that, in the absence of microwave field (figure 2-a), the steady state entropy becomes zero. By applying the microwave field to the lower levels (figure 2-b), for $\Omega_L \neq \Omega_R$ (dashed) the non-zero steady state entropy is established, while for $\Omega_L = \Omega_R$ (solid) the system becomes disentangled. Thus, the steady state DEM depends on the intensity of applied fields, and it can be controlled by intensity of either coherent optical or microwave fields.

Relative phase of applied fields is another important parameter to control the DEM in a closed-loop atomic system. In figure 3, we display time evolution of DEM for different values of relative phase of applied fields in the multi-photon resonance condition. Using parameters are $\Omega_m = 0.5\gamma$, $\Omega_L = 0.25\gamma$, $\Omega_R = 0.25\gamma$ (a), $0.5\gamma$ (b), $\Delta_L = \Delta_R = \Delta_m = 0.0\gamma$, $\varphi = 0.0$ (solid), $\varphi = \pi/4$ (dashed), $\varphi = \pi/2$ (dotted). It is shown that the steady state DEM depends on the relative phase of applied fields, and it can be controlled by either intensity or relative phase of applied fields. The interesting result is obtained for $\varphi = 0.0$ (solid) in which for $\Omega_L = \Omega_R$ (figure 3-a) the steady state entropy becomes zero while a nonzero entropy is obtained for $\Omega_L \neq \Omega_R$ (figure 3-b). Note that in the absence of microwave driving field, the closed-loop configuration is broken, and the DEM does not depend on the relative phase of applied fields.

To manifest phase dependent behavior of entanglement, in figure 4 we plot DEM versus relative phase of applied fields in the steady-state when $\gamma_{31} = \gamma_{32} = \gamma = 1$, $\Omega_m = 0.5\gamma$, $\Omega_L = 0.25\gamma$, $\Omega_R = 0.25\gamma$ (solid), $0.5\gamma$ (dashed), $\Delta_L = \Delta_R = 0$, $\Delta_m = 0$. One can see the entanglement shows periodic behavior versus relative phase of applied fields. The interesting behavior is obtained for $\Omega_R = \Omega_L$, in which the system is disentangled for $\varphi = n\pi$, $(n = 0, \pm 1, ...)$.

In figure 5, we display the steady state DEM versus the relative Rabi frequency of optical fields ($\Omega = \Omega_R / \Omega_L$) when other parameters are $\gamma_{31} = \gamma_{32} = \gamma = 1$, $\Delta_L = \Delta_R = \Delta_m = 0.0\gamma$, $\Omega_L = 0.25\gamma$, $\Omega_m = 0.5\gamma$, $\varphi = 0.0$ (solid), $\varphi = \pi/4$ (dashed) and $\varphi = \pi/2$ (dotted). It is found out that disentanglement just occurs for equal Rabi frequency of left and right fields in the zero relative phases of applied fields.

The physics of phenomena can be explained via the population distribution of dressed states. The dressed states of the system in the absence of microwave field are given by

$$|\varphi\rangle = \frac{1}{\sqrt{\Omega_L^2 + \Omega_R^2}}(\Omega_R|1\rangle - \Omega_L|2\rangle), \quad |\psi\rangle = \frac{1}{\sqrt{\Omega_L^2 + \Omega_R^2}}(\Omega^*_L|1\rangle + \Omega^*_R|2\rangle) \text{ and } |3\rangle. \quad (9)$$

By applying the microwave driving field to the system, the dressed states for $\varphi = 0$ and $\Omega_L = \Omega_R$ can be written as

$$|NC\rangle = \frac{1}{\sqrt{2}}(|1\rangle - |2\rangle), \quad |C\rangle = \frac{1}{\sqrt{2}}(|1\rangle + |2\rangle) \text{ and } |3\rangle \quad (10)$$

which according to equations (9), are same as the dressed states of the system in the absence of microwave field for $\Omega_L = \Omega_R = \Omega_0$.

The corresponding dressed state for $\varphi = \pi/2$ and $\Omega_L = \Omega_R = \Omega_0$ are given by

$$|A\rangle = \sqrt{\frac{\Omega_m^2}{\Omega_m^2 + 2\Omega_0^2}} \left( -i\frac{\Omega_0}{\Omega_m}|1\rangle + \frac{i\Omega_0}{\Omega_m}|2\rangle + |3\rangle \right),$$

$$|B\rangle = \sqrt{\frac{\Omega_0^2}{\Omega_m^2 + 2\Omega_0^2}} \times$$

$$\left( \left( -\frac{\sqrt{\Omega_m^2 + 2\Omega_0^2}}{\Omega_0} + \frac{\Omega_m^2 + \Omega_0^2}{\Omega_0(-i\Omega_m + \sqrt{\Omega_m^2 + 2\Omega_0^2})} \right)|1\rangle - \frac{\Omega_m^2 + \Omega_0^2}{\Omega_0(-i\Omega_m + \sqrt{\Omega_m^2 + 2\Omega_0^2})}|2\rangle + |3\rangle \right)$$

$$|C\rangle = \sqrt{\frac{\Omega_0^2}{\Omega_m^2 + 2\Omega_0^2}} \times$$

$$\left( \left( \frac{\sqrt{\Omega_m^2 + 2\Omega_0^2}}{\Omega_0} - \frac{\Omega_m^2 + \Omega_0^2}{\Omega_0(i\Omega_m + \sqrt{\Omega_m^2 + 2\Omega_0^2})} \right)|1\rangle + \frac{\Omega_m^2 + \Omega_0^2}{\Omega (i\Omega_m + \sqrt{\Omega_m^2 + 2\Omega_0^2})}|2\rangle + |3\rangle \right)$$

(12)

The dynamical behaviors of different dressed states population are shown in figure 6. Using parameters are $\Omega_m = 0(a,b), 0.5\gamma(c,d)$, $\Omega_L = 0.25\gamma$, $\Omega_R = 0.25\gamma(a,c,d)$, $0.5\gamma$ (b), $\varphi = 0(c)$, $\pi/2(d)$. Other parameters are same as in figure 2. The population of $|\phi\rangle$ (solid), $|\psi\rangle$ (dashed) and $|3\rangle$ (dotted) are plotted for $\Omega_L = \Omega_R$ (a) and $\Omega_L \neq \Omega_R$ (b). Note that $|\phi\rangle$ is a dark state and during interaction, the population of other two states will be transferred to this state. Thus, the steady state DEM of the system becomes zero (figure 2-a).

For $\Omega_L = \Omega_R$ all of population are equally distributed in two lower levels according to the coherent population trapping and applying the microwave driving field to the lower levels cannot affect the population distribution of the system. Then the same results are obtained for $\Omega_L = \Omega_R$ and $\varphi = 0$ in the absence (figure 6-a) or presence (figure 6-c) of microwave driving field, and all of population is transferred to dark state. The situation is completely different for $\varphi = \pi/2$ in which, all of three dressed states are populated (figure 6-d) and subsequently the non-zero steady state entropy is obtained as shown in dotted line of figure 3-a. Thus the atom-photon entanglement just occurs when the population of the system does not trap only in one of the dressed states of the system.

To compare our numerical results with analytical calculation, in figure 7, we display the DEM versus relative phase and relative Rabi frequency of optical fields. Using parameters are $\Omega_L = 0.25\gamma$, $\Omega_m = 0.5\gamma$, $\Delta_L = \Delta_R = \Delta_m = 0.0\gamma$, $\gamma_{31} = \gamma_{32} = \gamma = 1$. It is shown that the analytical solutions are in good agreement with the numerical results. The period of DEM variation is $\varphi = \pi$ which can be explained via obtained analytical results.

Now, we are interested to investigate the time-dependent behavior of quantum entropy beyond the multi-photon resonance condition ($\Delta \neq 0$). In figure 8, we plot the DEM for different relative phase of applied fields, beyond the multi-photon resonance condition for $\gamma_{31} = \gamma_{32} = \gamma = 1$, $\Delta_L = 0.5$, $\Delta_R = \Delta_m = 0.0\gamma$, $\Delta = 0.5\gamma$, $\Omega_L = 0.25\gamma, \Omega_R = 0.5\gamma$, $\Omega_m = 0.5\gamma$, $\varphi = 0.0$ (solid), $\varphi = \pi/4$ (dashed), $\varphi = \pi/2$ (dotted). An investigation on figure 8 shows that the DEM does not get steady state solution, beyond multi-photon resonance condition.

Finally we are interested to take into account the effect of Doppler effect on the obtained results. Doppler broadening due to the atom's thermal velocity $\vec{v}$ can be applied by replacing the $\Delta_L$, $\Delta_R$ and $\Delta_m$ by $\Delta_L - \vec{k}_L.\vec{v}$, $\Delta_R - \vec{k}_R.\vec{v}$ and $\Delta_m - \vec{k}_m.\vec{v}$, respectively. The parameters $\vec{k}_L$, $\vec{k}_R$ and $\vec{k}_m$ are the wave vectors of left, right and microwave driving fields. To obtain the steady state solutions for set of equations (2), it is assumed that the multi-photon resonance condition, i. e. $\Delta(\vec{v}) = \Delta_L - \Delta_R - \Delta_m - \vec{k}_L.\vec{v} + \vec{k}_R.\vec{v} + \vec{k}_m.\vec{v} = 0$ is still satisfied by detunings and propagation directions of applied fields. It is worth noting that a suitable choice of laser propagation directions allows us to obtain the steady state results even in the presence of Doppler effect. The Maxwellian velocity distribution,

$$f(v) = \frac{1}{\sqrt{2\pi}u}\exp[-v^2/u^2], \tag{13}$$

is used to average our results for populations and quantum entropy. The parameter $u = \sqrt{2k_B T/m}$ shows the Doppler width of distribution; $m$ is mass of moving atom, $T$ is temperature of the cell and $k_B$ is Boltzmann constant. Then the Doppler-averaged quantum entropy of the system is given by

$$\bar{S} = \int_{-\infty}^{\infty} f(v)S(v)dv.$$

Figure 9 shows the steady state DEM versus relative phase of applied fields in the presence of Doppler effect for $ku = 0.5\gamma$ (solid), $5\gamma$ (dashed), $20\gamma$ (dotted). We assume $|\vec{k}_L| = |\vec{k}_R| = k$ and other using parameters are same as the dashed line of figure 4. It can be seen that increasing the Doppler width of distribution leads to reduction of the maximum values of Doppler-averaged quantum entropy and it has a destructive role in atom-photon entanglement.

**Conclusions**

In conclusion, we investigated the atom-photon entanglement in a three-level Λ-type closed-loop atomic system via the von Neumann entropy. The results were obtained in multi-photon resonance condition and beyond it. It was found that the DEM in such a system is phase-dependent and the nonzero steady state entropy was obtained for different values of relative phase of applied fields. We included the Doppler broadening due to the thermal motion of atoms and it was found that increasing the Doppler width of distribution decrease the steady state DEM.

**Figures caption**

**Figure 1.** A three-level $\Lambda$-type closed-loop atomic system driven by two coherent optical fields and a microwave field with corresponding Rabi frequencies $\Omega_L$, $\Omega_R$ and $\Omega_m$, respectively.

**Figure 2.** Dynamical behavior of the reduced atomic entropy via different values of the intensity of applied fields in multi-photon resonance condition ($\Delta = 0$). The selected parameters are $\Delta_L = \Delta_R = \Delta_m = 0.0\gamma$, $\Omega_L = 0.25\gamma$, $\Omega_R = 0.25\gamma$ (solid), $0.5\gamma$ (dashed), $\Omega_m = 0$ (a), $0.5\gamma$ (b).

**Figure 3.** Time evolution of DEM for different relative phase of applied fields in multi-photon resonance condition. Using parameters are $\Omega_m = 0.5\gamma$, $\Omega_L = 0.25\gamma$, $\Omega_R = 0.25\gamma$ (a), $0.5\gamma$ (b), $\Delta_L = \Delta_R = \Delta_m = 0.0\gamma$, $\varphi = 0.0$ (solid), $\varphi = \pi/4$ (dashed), $\varphi = \pi/2$ (dotted).

**Figure 4.** The steady state DEM versus relative phase of applied fields for $\gamma_{31} = \gamma_{32} = \gamma = 1$, $\Delta_L = \Delta_R = 0$, $\Delta_m = 0$, $\Omega_m = 0.5\gamma$, $\Omega_L = 0.25\gamma$, $\Omega_R = 0.25\gamma$ (solid), $0.5\gamma$ (dashed).

**Figure 5.** The steady state DEM versus the relative Rabi frequency of optical fields ($\Omega = \Omega_R/\Omega_L$). Using parameters are $\gamma_{31} = \gamma_{32} = \gamma = 1$, $\Delta_L = \Delta_R = \Delta_m = 0.0\gamma$, $\Omega_L = 0.25\gamma$, $\Omega_m = 0.5\gamma$, $\varphi = 0.0$ (solid), $\varphi = \pi/4$ (dashed) and $\varphi = \pi/2$ (dotted).

**Figure 6.** The dynamical behavior of different dressed states population. Using parameters are $\Omega_m = 0(a,b), 0.5\gamma(c,d)$, $\Omega_L = 0.25\gamma$, $\Omega_R = 0.25\gamma(a,c,d)$, $0.5\gamma$ (b), $\varphi = 0(c)$, $\pi/2(d)$. Other parameters are same as in figure (2). The population of $|\phi\rangle$ (solid), $|\psi\rangle$ (dashed) and $|3\rangle$ (dotted) are plot for $\Omega_L = \Omega_R$ (a) and $\Omega_L \neq \Omega_R$ (b).

**Figure 7.** Three dimensional plot of atomic entropy versus relative Rabi frequency and relative phase of applied fields. The selected parameters are $\Delta_L = \Delta_R = \Delta_m = 0.0\gamma$, $\gamma_{31} = \gamma_{32} = \gamma = 1$, $\Omega_m = 0.5\gamma$, $\Omega_L = 0.25\gamma$.

**Figure 8.** DEM for different relative phase of applied fields, beyond the multi-photon resonance condition for $\gamma_{31} = \gamma_{32} = \gamma = 1$, $\Delta_L = 0.5$, $\Delta_R = \Delta_m = 0.0\gamma$, $\Delta = 0.5\gamma$, $\Omega_m = 0.5\gamma$, $\Omega_L = 0.25\gamma$, $\Omega_R = 0.5\gamma$, $\varphi = 0.0$ (solid), $\varphi = \pi/4$ (dashed), $\varphi = \pi/2$ (dotted).

**Figure 9.** The steady state DEM versus relative phase of applied fields in the presence of Doppler effect for $ku = 0.5\gamma$ (solid), $5\gamma$ (dashed), $20\gamma$ (dotted). We assume $|\vec{k}_L| = |\vec{k}_R| = k$ and other using parameters are same as the dashed line of figure 4.

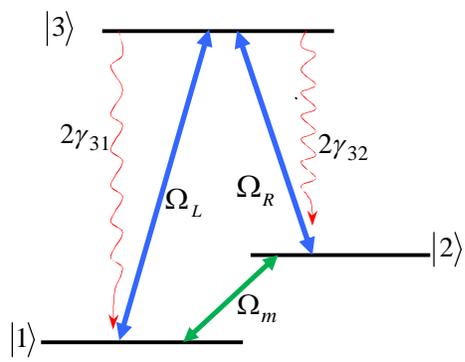

Figure 1

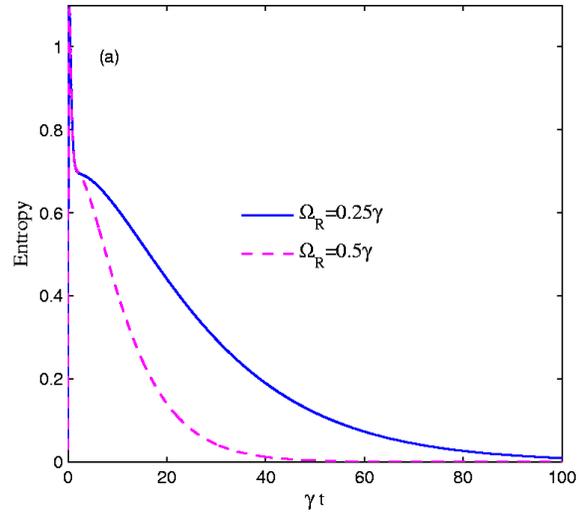

Figure 2-a

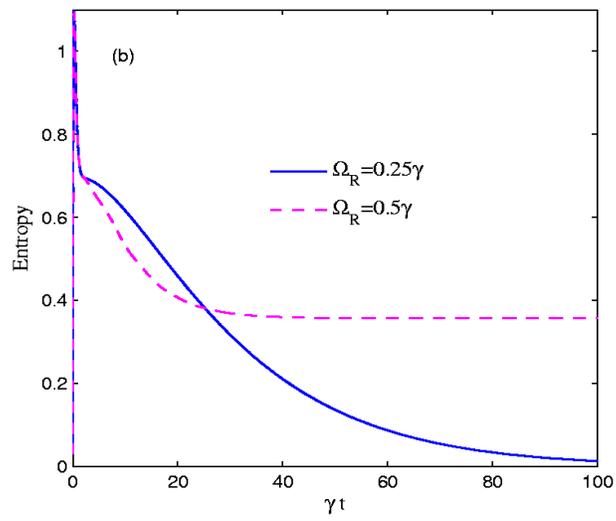

Figure 2-b

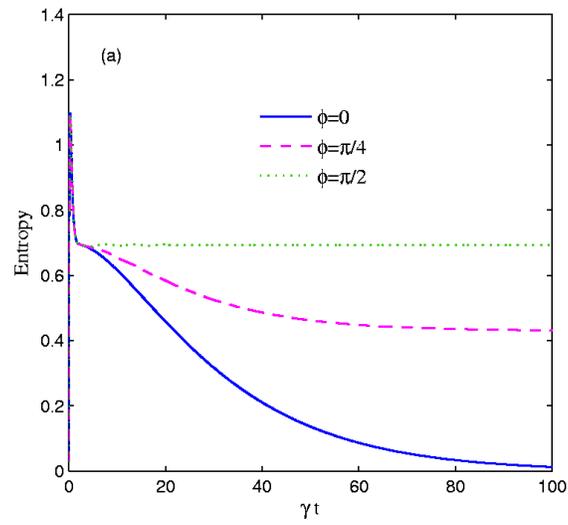

Figure 3-a

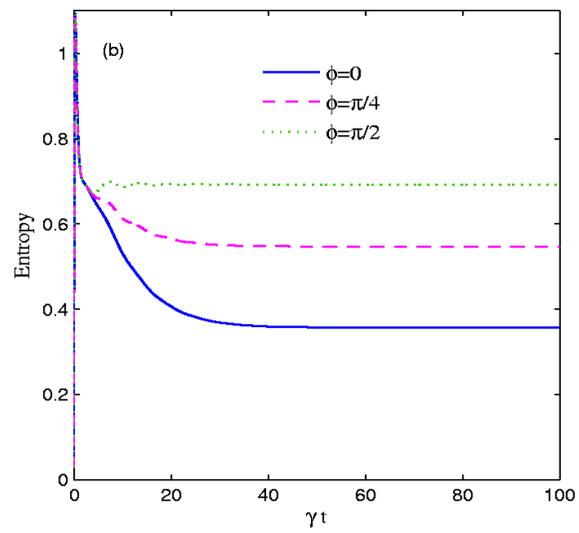

Figure 3-b

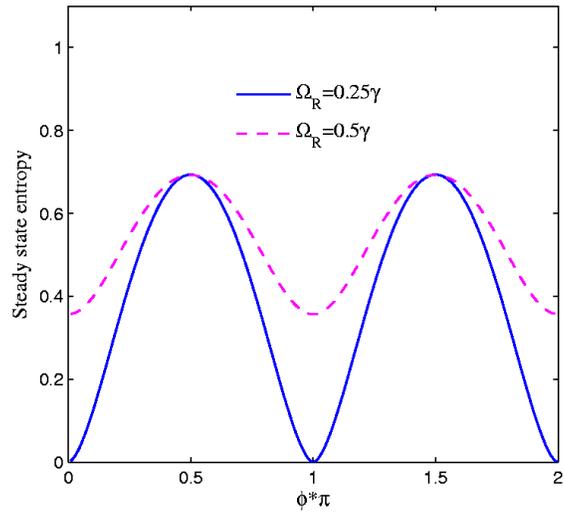

Figure 4

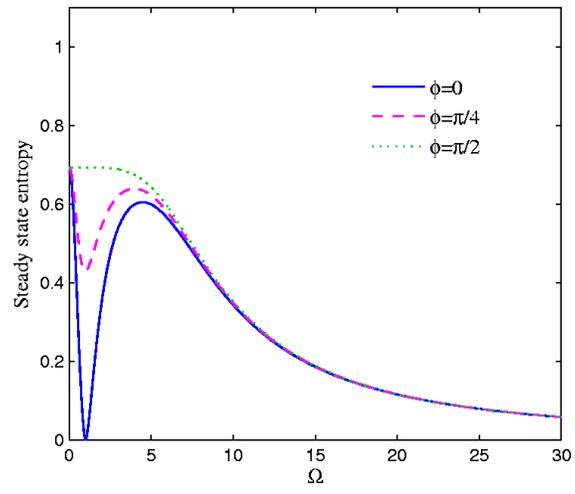

Figure 5

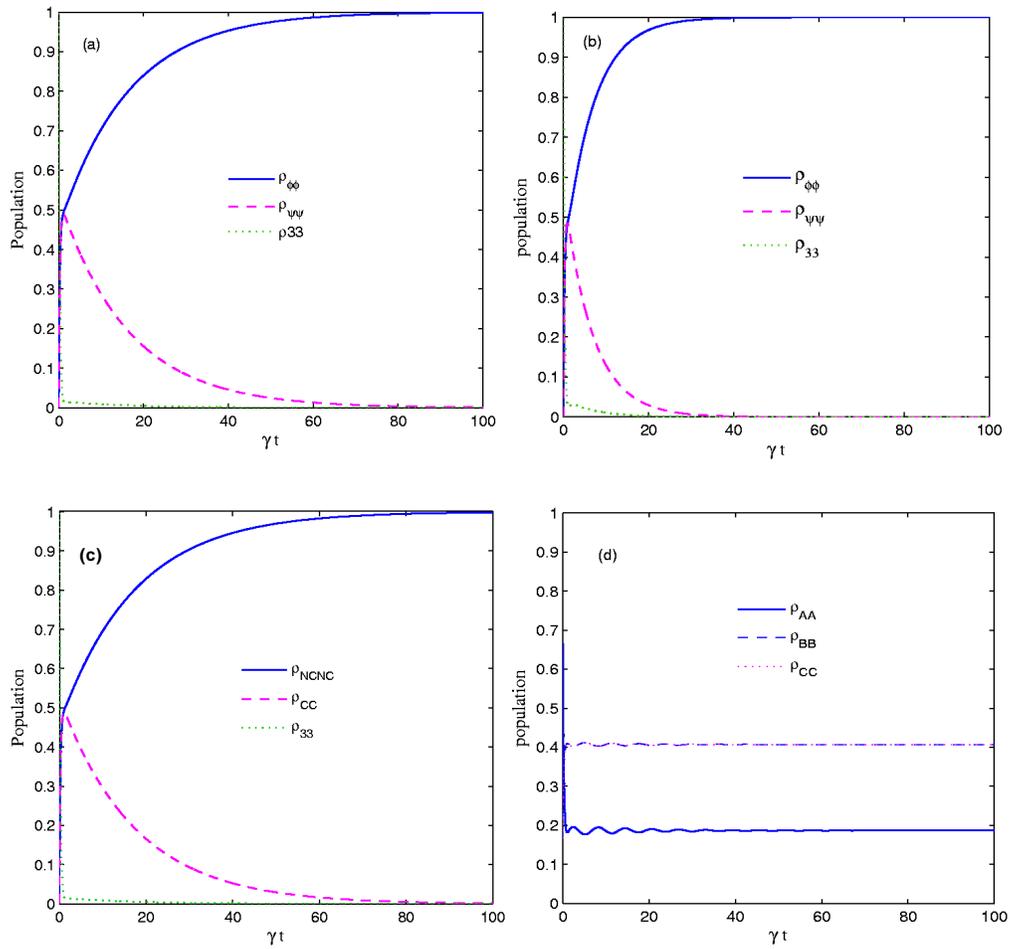

Figure 6

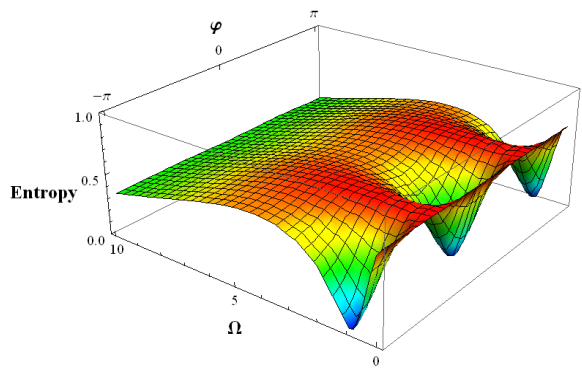

Figure7

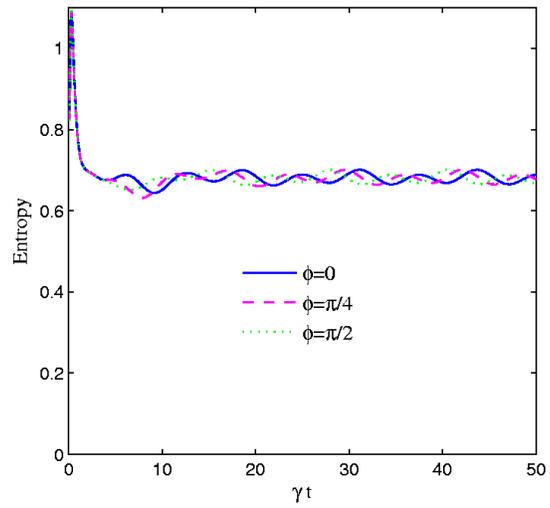

Figure 8

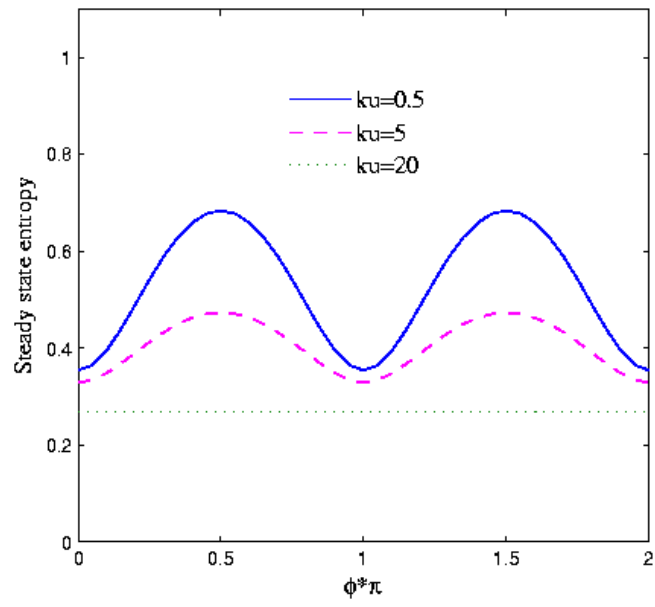

Figure 9